\documentclass[preprint2]{aastex}

\usepackage{natbib}
\usepackage{graphicx}
\usepackage{txfonts}

\begin{document}

\shorttitle{X-ray Properties of MAXI J1535-571}
\shortauthors{Shang et al.}
\title{Evolution of X-ray Properties of MAXI J1535-571: Analysis With The TCAF Solution} 
\author{J.-R. Shang\altaffilmark{1}, D. Debnath\altaffilmark{2*}, D. Chatterjee\altaffilmark{2}, A. Jana\altaffilmark{2}, S. K. Chakrabarti\altaffilmark{3,2}, H.-K. Chang\altaffilmark{1,4}, 
Y.-X. Yap\altaffilmark{1}, C.-L. Chiu\altaffilmark{1}}
\altaffiltext{1}{Institute of Astronomy, National Tsing Hua University, Hsinchu 30013, Taiwan}
\altaffiltext{2}{Indian Centre for Space Physics, 43 Chalantika, Garia St. Rd., Kolkata, 700084, India}
\altaffiltext{3}{S. N. Bose National Centre for Basic Sciences, Salt Lake, Kolkata, 700106, India}
\altaffiltext{4}{Department of Physics, National Tsing Hua University, Hsinchu 30013, Taiwan}
\email{dipakcsp@gmail.com}


\begin{abstract}
We present spectral and timing properties of the newly discovered X-ray transient source, MAXI J1535-571, 
which is believed to be a Galactic X-ray binary containing a black hole candidate (BHC) as the primary object. 
After its discovery on 2017 Sep. 2, it has been monitored regularly in multi-wavelength bands by several 
satellites. We use archival data of Swift (XRT and BAT) and MAXI (GSC) satellite instruments to study 
accretion flow dynamics of the source during the outburst. During its outburst, the source became very bright 
in the sky with a maximum observed flux of $5$~Crab in the $2-10$~keV GSC band. Similar to other transient BHCs, 
it also shows signatures of low frequency quasi-periodic oscillations (QPOs) during the outburst. Spectral data 
of different instruments are fitted with the transonic flow solution based two-component advective flow (TCAF) 
model fits file to find the direct accretion flow parameters. Evolution of spectral states and their transitions 
are understood from the model fitted physical flow parameters and nature of QPOs.
We also estimate probable mass of the black hole from our spectral analysis as $7.9-9.9~M_\odot$ or $8.9\pm1.0~M_\odot$. 
\end{abstract}


\section{Introduction}

MAXI J1535-571 is a new Galactic X-ray binary. It is detected on 2017 September 2 simultaneously by MAXI/GSC (Negoro et al. 2017a) 
and Swift/BAT (Kennea et al. 2017) at the sky location of R.A.$=15^h 35^m 10^s $, Dec$=-57^\circ 10' 43''$ (Negoro et al. 2017a,b) near 
the Galactic plane. After its discovery, this source has been monitored and studied extensively in multi-wavelength bands, i.e., in radio 
(Russell et al. 2017a; Tetarenko et al. 2017), optical (Scaringi et al. 2017), near-infared (Dincer 2017), X-ray (Negoro et al. 2017a; 
Kennea et al. 2017; Shidatsu et al. 2017a,b; Xu et al. 2017) and $\gamma$-ray (Britt et al. 2017). The optical and near-infared counterpart 
of the source is reported by Scaringi et al. (2017) and Dincer (2017) respectively. Nakahira et al. (2017) have reported spectral softening 
on 2017 Sep. 10 from MAXI/GSC analysis and Kennea (2017) has reported source brightening and softening on 2017 Sep. 11 from Swift/XRT analysis. 
Mereminskiy and Grebenev (2017) have observed a sharp quasi-periodic oscillation (QPO) of $1.9$~Hz frequency on 2017 Sep. 11. 
Radio observations by Russell et al. (2017a) and Tetarenko et al. (2017) suggest that source had active jets during its hard-state 
of the outburst. Xu et al. (2017) have predicted the source as a high spinning object with spin parameter $a > 0.84$. 
Miller et al. (2018) have estimated the spin of the source as $a=0.994$ using NICER data.
Shidatsu et al. (2017a) have reported that after Sep. 10, 2017, further softening occurred in the source spectrum and on 2017 Sep. 19 
the source was in the intermediate state and very likely to enter the soft state. Russell et al., (2017b) have observed a radio 
re-brightening on 2017 Oct. 25. They also have reported a decreasing QPO frequency on 2017 Oct. 11, 22 and 24, suggesting the source 
transiting back in towards hard state. After this sudden spectral hardening and a decrease in flux around 2017 Oct. 13, MAXI J1535-571
shows consistent spectral softening since 2017 Oct. 25 (Shidatsu et al. 2017b). Around late November (2017 Nov. 23-25) the source 
shows the softest spectrum and may be in the high/soft state (Shidatsu et al., 2017b). Nakahira et al. (2018) have reported an almost
exponential decrease in flux of MAXI J1535-571 after around (2018 Jan. 12; MJD=58130) while it is still in the soft state (Negoro et al., 2018).
The softening continued and around 2018 Apr. 16 the flux decreased below the detection limit of MAXI/GSC and it remained there for 
$\sim 4$ days. Swift/XRT observation on 2018 Apr. 26, 30 and May 2 shows a decrease in soft X-ray flux and increase in hard X-ray 
emission (Russell et al., 2018). Parikh et al. (2018) have suggested that the source is in the hard state on 2018 May 14 by spectral 
fitting the Swift/XRT data considering a fixed hydrogen column density of $4.4\times10^{22}~cm^{-2}$. The spectral fitting was done 
with an absorbed power-law model and the obtained photon index is $\Gamma\sim1.7$. A radio detection (with an index of $\alpha\sim0.3$) 
is also observed on 2018 May 14 by Australia Telescope Compact Array (ATCA), consistent with an compact X-ray jet emission observed 
in the hard state (Parikh et al., 2018). They also fitted XRT spectra from 2018 May 18 with an absorbed power-law model with $n_H$ 
fixed at $4.4\times10^{22}~cm^{-2}$ and obtained $\Gamma\sim4.7$. Further, quasi simultaneous ATCA and ALMA (Atacama Large 
Millimeter/sub-millimeter Array) did not detect the source on 2018 May 17 implying that the radio jet observed on 2018 May 14 might 
have quenched, generally a signature of the soft state. Recent MAXI/GSC observation indicates that MAXI J1535-571 started to go through 
another state transition (soft to hard) around 2018 May 26 (Negoro et al., 2018b). The spectral and temporal nature of 
the source as have been reported in the literature indicates that MAXI~J1535-571 is a potential low-mass X-ray binary (Scaringi et al. 2017) 
and more specifically it is a black hole (BH) binary system (Negoro et al. 2017b). 
Tao et al. (2018) have recently analyzed the spectral properties of the source using a phenomenological 
disk blackbody plus power-law models. 

Transient black hole candidates (BHCs) generally show rapid evolutions of their spectral and temporal properties during an outburst. 
In an outburst, classical or type-I sources show four different spectral states: Hard (HS), Hard-Intermediate (HIMS), Soft-Intermediate (SIMS) 
and Soft (SS), which form a hysteresis loop in the sequence : HS$\rightarrow$HIMS$\rightarrow$SIMS$\rightarrow$SS$\rightarrow$SIMS 
$\rightarrow$HIMS$\rightarrow$HS (see Debnath et al. 2013, 2017 and references therein). Observation of low and high frequency QPOs 
are characteristic temporal feature of some of these spectral states. Depending upon the nature (centroid frequency, width, Q-value, 
rms amplitude, etc.) these low frequency QPOs are divided into three types: A, B, C. 
Although there is significant debate around the nature of the QPO, here we shall assume that they are originated
due to the oscillation of Compton cloud and will discuss our results in the framework of CENBOL i.e., shock oscillation 
model (Molteni et al. 1996; Ryu et al. 1997; Chakrabarti et al. 2015). Here, due to resonance, the observed QPO 
frequencies are inversely proportional to the infall time scales. Type-C QPOs observed in HS \& HIMS are mainly due to 
resonance oscillation, when cooling time scale roughly matches with the infall time scale. QPOs may also occur due to 
non-satisfaction of Rankine-Hugoniot condition which is required to form a stable shock (Ryu et al., 1997; Chakrabarti et al., 2015) 
or due to weak resonance phenomenon (either a weakly resonating CENBOL for type `B' or a shockless centrifugal barrier for type `A'). 

MAXI~J1535-571 is an interesting source to study in X-rays, since it has showed rapid evolution of its temporal and spectral properties 
during the short initial period of its outburst phase in 2017. Studying spectral data with the phenomenological model such as, combined 
disk black body (DBB) and power-law (PL) model gives us a general overview of thermal and non-thermal radiation fluxes. 
The thermal component in the form of multi-color blackbody radiation is contributed from the standard Keplerian disk (Shakura \& Sunyaev 1973) 
and the power-law, i.e., the hard component mainly originates from a `Compton" cloud (Sunyaev \& Titarchuk 1980, 1985) 
consisting of hot electrons, whose thermal energy is transferred to low-energy photons emitted from the Keplerian disk by 
repeated Compton scatterings to produce hard X-ray photons. 
Two component advective flow (TCAF) model (Chakrabarti \& Titarchuk 1995; Chakrabarti 1997) comes from the solution of
radiative hydrodynamics equations, where `hot' Compton cloud of earlier models has been replaced by the CENtrifugal pressure supported
BOundary Layer (CENBOL), which automatically forms behind the centrifugal barrier due to piling up of the low viscous (sub-critical),
low angular momentum and optically thin matter known as sub-Keplerian or halo accretion component of the flow. This is really a centrifugal
pressure supported shock wave which may also oscillate under suitable conditions decided by the flow parameters. The other component 
of accretion flow is highly viscous, high angular momentum, geometrically thin and optically thick Keplerian or disk matter
which is submerged inside the halo component. According to this model, a Keplerian disk is naturally truncated at the shock location, 
which is the outer boundary of the CENBOL. This Keplerian disk component settles down to a standard Shakura \& Sunyaev (1973) 
disk, when cooling is efficient. Low energy (soft) thermal photons from the Keplerian disk intercept with the CENBOL 
(composed of hot electrons) and emit as high energy (hard) photons by cooling down CENBOL through repeated inverse-Compton scattering.
Some part of the emitted hard photons reflect from the Keplerian disk locally changing its temperature. The
radiations re-emitted by this modified disk of slightly higher temperature interacts with CENBOL. This 
iterative process creates a so-called reflection component self-consistently. 

Recently, after implementation of this TCAF solution into HEASARC's spectral analysis software package XSPEC (Arnuad 1996) 
as an additive table model 
(Debnath et al. 2014, 2015a), it has been found that TCAF model can successfully extract accretion flow parameters of many BHCs 
during their X-ray active phases (Debnath et al. 2014; 2015a,b; 2017; Mondal et al. 2014, 2016; Jana et al. 2016; 
Chatterjee et al. 2016, 2019; Molla et al. 2017; Bhattacherjee et al. 2017).
Estimation of intrinsic source parameters, such as mass, spin, etc. have been done quite successfully from TCAF model fitted 
spectral analysis (see, Molla et al. 2016, 2017; Debnath et al. 2017). 
These results motivated us to study 2017-18 outburst of MAXI~J1535-571 with the TCAF model.

The paper is organized in the following way. In \S 2, we briefly discuss observations and data analysis procedure with the TCAF 
model {\it fits} file. In \S 3, we present timing and spectral analysis result with archival data of Swift and MAXI satellites. 
We also estimate the mass of the BH object MAXI~J1535-571 from TCAF model spectral fit and photon index-QPO frequency 
scaling method (Shaposhnikov \& Titarchuk 2007).
Finally in \S 4, a brief discussion and concluding remarks based on our study of the source are presented. 

\section{Observations and Data Analysis}

We used archival data taken by Swift/XRT combined (where available) with Swift/BAT and MAXI/GSC 
to study the spectral and timing properties for MAXI J1535-571 during its 2017-18 outburst. 
To have extended harder energy band of the spectra, we combined XRT spectra with either BAT and/or GSC spectra of the same day. 
MAXI~J1535-571 spectra are analyzed roughly on a daily basis from 2017 Sep 4 (MJD=58000.71) to 2017 Oct 24 (MJD=58050.98) with 
two types of models: $i)$ PL model or DBB+PL models, and $ii)$ TCAF model {\it fits} file 
in HEASARC's spectral analysis software package XSPEC version 12.9.0. 
Interstellar absorption model {\it `Tbabs'} was used while fitting spectra and its parameter hydrogen column density ($N_{H}$) was kept free.
We used `wilm' abundances and `vern' scattering cross-section for our spectral analysis. No systematic error is used during 
the entire period of our analysis. All the data we used in this work are listed in Tables 1 \& 2. 
The source was discovered on Sep 2, 2017 but due to the poor S/N ratio, and to avoid XRT pileup problem, we have not used the 
initial four observations (IDs: 00770431000, 00770431001, 00770502000, \& 00770656000) of the outburst. 
In observation ID: 00010264011 of Sep 19, the source was shifted 
to the edge of the FOV and failed the spectrum data reduction pipeline, so we abandoned this pointing as well. 
We also abandoned observation ID: 00088245004 
of Sep. 22 due to observation of two bright sources within 30 arc-sec FOV of the source. 

Swift/XRT spectra are reduced using Swift web tool provided by UK Swift Science Data Centre at the University of Leicester (Evans et al. 2009). 
MAXI/GSC spectra are produced by MAXI on-demand process web tool (Matsuoka et al. 2009). Swift/BAT data are reduced through the standard pipeline. 
TCAF model requires four mandatory and two auxiliary input parameters:
(i) the Keplerian/disk accretion rate ($\dot{m}_d$ in units of $\dot{M}_{Edd}$), (ii) the sub-Keplerian/halo accretion 
rate ($\dot{m}_h$ in units of $\dot{M}_{Edd}$), (iii) the location of the shock ($X_s$ in units of Schwarzschild radius $r_s = \frac{2GM_{BH}}{c^2}$), 
(iv) the compression ratio $R =\frac{\rho_+}{\rho_-}$, where $\rho_+$ and $\rho_-$ are densities of the post- and pre-shock matters, (v) the mass of the 
black hole ($M_{BH}$ in units of $M_\odot$), 
and (vi) the model normalization value ($N$). Normally, if the mass is already known and the model normalization (scaling between observed and
calculated spectra) is already calculated for the source and the instruments, then the remaining four mandatory parameters would suffice to fit a
spectrum. Otherwise, if the model normalization is kept in a narrow range, the mass also comes out independently from each observation. 
The solution table was read by XSPEC and corrected with built-in interstellar absorption model as $Tbabs \times TCAF.fits$, while fitting spectra. 
For timing analysis to study QPO behaviour, we used POWSPEC 1.0 (XRONOS 5.22) package on all available XRT data from 2017 Sep 2 to 2017 Oct. 24. 
MAXI~J1535-571 was one of the brightest X-ray sources in the sky during its outburst. So the light curves are only processed by GTI subtraction 
and barycentric time correction within XRT 23.6 $\times$ 23.6 arcmin FOV. 
To find centroid frequency of the dominating primary QPOs, we fitted fast Fourier transformed 
(using `powspec' task) power density spectra (PDS) of the $0.01$~sec binned XRT light curves with Lorentzian profiles.

\section{Results}

We studied both the temporal and the spectral properties of MAXI~J1535-571 during its initial $\sim 2$~months of the 2017-18 outburst 
to find physical picture about the accretion flow properties of the source. We used archival and on-demand data of the Swift XRT \& BAT 
and the MAXI/GSC instruments in between 2017 Sep. 2 (MJD=57998.82) to 2017 Oct. 24 (MJD=58050.98) for our study. 
The results based on our analysis are presented in the following sub-Sections.

\subsection{Evolution of Timing Properties}

We studied daily variations of the photon count rates of the source during the outburst using MAXI/GSC, Swift/BAT data from beginning 
of the outburst to 2018 June 3 (MJD=58272). In Fig. 1a, GSC and BAT one day average photon fluxes (in units of mCrab) 
in $2-10$~keV and $15-50$~keV energy bands respectively are shown. During its outburst, it became one of the brightest X-ray sources in the sky. 
The maximum flux of MAXI~J1535-571 became about $5$~Crab in $2-10$~keV GSC band, which is much higher than that in the quiescent state.
To draw an outline about the spectral states during the outburst, we also studied variation of the hardness ratio (HR). 
The variation of HRs, which are defined as ratio between BAT with GSC fluxes, and is shown in Fig. 1b.  
In outburst profiles (as in Fig. 1a), the count rate rises rapidly during the rising phase of the outburst. During the declining 
phase of the outburst opposite scenario with a slower decay is observed. According to Debnath et al. (2010), current outburst of 
MAXI~J1535-571 could be termed as fast-rise-slow-decay (FRSD) type. The period of our spectral analysis (from Sep. 4 to Oct. 24, 2017) 
are marked by the region between two vertical lines.

In Fig. 1 of Nandi et al. (2012) a good example is presented using the 2010-11 outburst of GX~339-4 to define spectral states based on HRs.
Here, during the current outburst of MAXI~J1535-571, BAT triggered 3 days after (on Sep. 5, 2017) the discovery of the source. 
Slow decay in HR from $2.83$ to $2.29$ in between Sep. 5-7, 2017 (MJD=58001-58003) indicated that during this period, the source was in HS. 
After that, a rapid fall in HR from $2.29$ to $0.88$ in the next five days (Sep. 7-11, 2017; MJD=58003-58007) 
indicated that the source could be in HIMS during this period of the outburst. After that, the HR showed 
sporadic behaviour while remaining low till 2017 Nov. 18 (MJD=58075). At this point it reached a very low value 
of $\sim 0.004$. This phase of the outburst could be termed as SIMS. After that, the HR remained at its low values 
till 2018 Apr 2 (MJD=58210), when a rise in HR ($=0.35$) was observed. This period of the outburst source could be termed as SS. 
After 2018 Apr. 2, high fluctuations in HRs in between $0.01-1.86$ were observed, which indicated that the source probably moved again 
in hard or intermediate spectral states of the declining phase of the outburst. A similar report of the declining harder spectral states 
is found in by Parikh et al. (2018) who suggested that it occurred on May 14 (MJD=58252).

Low frequency ($\sim 0.01-30$~Hz) QPOs are very important characteristic temporal feature of stellar massive BHCs. 
The QPOs are strongly correlated with observed spectral states. 
To find QPOs during the current outburst of MAXI~J1535-571, Fourier transformed PDS with time binned $0.01$s Swift/XRT 
light curves in energy range 1-10 keV were analyzed. Leahy-normalized PDS fitted with Lorentzian profiles. The fitting parameters 
(centroid frequency, FWHM, Q-value, rms amplitude, etc.) are listed in Table 1. QPOs of frequencies $0.44-6.48$~Hz were observed 
sporadically (on and off) in $15$~observations out of total $27$~observation IDs.
Based on the classification criteria as mentioned in Motta et al (2011), we found two classical type (A \& C) and one 
unknown type (marked as type-X) of QPOs during the outburst. In Fig. 2, we show PDS of these three different type of QPOs 
fitted with Lorentzian models. In Fig. 2a \& b, type-A QPO for observation ID: 00010264023 (observed on 2017 Oct. 3; MJD=58029.73), 
and type-C QPO for observation ID: 00010264009 (MJD=58012.11; UT date: 2017 Sep. 16) are shown respectively. The $0.44$~Hz QPO 
of observation ID: 00010264003 (MJD=58004.27; UT date: 2017 Sep. 8) is shown in Fig. 2c, does not belong to the normal type of QPOs. 
However it looked like a C type consisting of a break frequency. However, lesser Q-value (=$2.34\pm0.81$) and 
higher rms (=$8.59\%\pm1.56$) did not allowed us to define it as a type-C. So, we marked it as unknown type-X.

\subsection{Evolution of Spectral Properties}

Combined Swift/XRT, Swift/BAT, MAXI/GSC (whenever BAT and/or GSC data are available) or only Swift/XRT spectral data of 27 observations 
of MAXI~J1535-571 during the initial outburst period (from 2017 Sep. 4 to 2017 Oct. 24) are analyzed to understand accretion flow properties of 
the source during its very first outburst after the discovery. To know about the gross nature of the variation of the photon fluxes, which are 
produced by the thermal and the non-thermal processes, we first fitted spectra with the phenomenological PL model or DBB+PL models. 
After that, to find physical picture about the accretion flow dynamics of the source during the outburst, spectral data are fitted with the 
latest version (v0.3) of the TCAF model {\it fits} file as an additive table model in XSPEC. 
In Table 2, we mention TCAF model fitted or derived 
parameters in details. To get an idea about the nature of the spectral states, we put values of the PL model fitted photon 
indices in Col. 4 of Table 2.

In Fig. 3, we show TCAF model fitted combined XRT ($1-9~keV$), GSC ($7-20$~keV), and BAT ($15-50$~keV) spectra of observation ID: 0001026010 
(MJD=58014.17; UT date: 2017 Sep. 18). 
In Fig. 4, variation of the accretion flow parameters ($\dot{m}_d$, $\dot{m}_h$, $X_s$, $R$) are shown. While in Fig. 5 (a-b), 
variation of the mass of the BH (since it was kept free while fitting spectra), derived accretion rate ratio (ARR=$\dot{m}_h$/$\dot{m}_d$) 
are shown. In Fig. 5(c-d), variation of the observed frequency of the dominating QPOs and phenomenological model fitted PL photon indices are shown.

Although daily data is not available, depending upon the nature of the variation of accretion flow parameters and observed QPOs, 
we can classify our studied period of the outburst into three spectral states: HS, HIMS and SIMS. The evolution of the spectral and the temporal 
properties during these observed states are mentioned below.

\subsubsection{Hard State (HS):}
According to the results of our analysis, first two observations on 2017 Sep. 4 \& 6 belong to the hard spectral state, 
since in these two observations, accretion flows are highly dominated by the sub-Keplerian halo component ($\dot{m}_h$=$1.38$, 
$1.62$~$\dot{M}_{Edd}$) as compared to the Keplerian disk component ($\dot{m}_d$=$0.16$, $0.18$~$\dot{M}_{Edd}$). 
For this reason, higher ARRs are observed. Relatively stronger shocks ($R=1.38, 1.23$) are observed far away ($X_s=269, 241$~$r_s$) 
from the BH horizon. Low values of PL photon indices ($\Gamma=1.39, 1.43$) and higher values of HR ($>2.0$) also support our 
placement of these two observations. Strangely, no monotonically evolving QPOs are found during these two observations, 
probably due to the non-satisfaction of resonance condition.

\subsubsection{Hard-Intermediate State (HIMS):}
Third observation on 2017 Sep. 8 (MJD=58004.27) belongs to this spectral state, since on this observation we see decrease 
in $\dot{m}_h$ (from $1.62$~$\dot{M}_{Edd}$ to $0.79$~$\dot{M}_{Edd}$) and increase in $\dot{m}_d$ (from $0.18$~$\dot{M}_{Edd}$ 
to $0.59$~$\dot{M}_{Edd}$). As a result of this, we see a sharp fall in ARR value at $\sim 1.34$ from its previous observed 
value of $\sim 9.0$. A unknown type QPO of $0.41$~Hz is present in this observation. The shock is found to move closer to the 
BH ($X_s=123$~$r_s$) with weaker strength. Since the ARR decreases rapidly due to fall in $\dot{m}_h$ and rise in $\dot{m}_d$ 
and the shock moved inward, we define this particular observation of MAXI~J1535-571 on 2017 Sep. 8 as HIMS. 
We also observe a slight increase in $\Gamma$ value ($\sim 1.47$) and decrease in HR ($\sim 1.96$) on this observation. 

\subsubsection{Soft-Intermediate State (SIMS):}
Starting from the fourth observation on 2017 Sep. 11 (MJD=58007.27) until the end of our detailed study (2017 Oct. 24), the source has 
moved to softer spectral state as the Keplerian disk rate started to dominate over the sub-Keplerian halo rate. QPOs are observed sporadically 
on and off in between $1.85-6.48$~Hz, which is a characteristic property of the SIMS (see, Nandi et al. 2012 and references therein). 
On the first observation of this spectral state, shock is moved close to the BH ($X_s=25$~$r_s$) with a weaker strength ($R=1.07$). 
On this particular observation, ARR and HR are reduced to low values of $\sim 0.50$, $\sim 0.88$ respectively and PL photon index is 
increased to $\sim 2.04$. 

Overall during this spectral state, two component accretion rates: $\dot{m}_d$ and $\dot{m}_h$ are varied in between 
$\sim 1.11-4.48$~$\dot{M}_{Edd}$, and $\sim 0.55-1.57$~$\dot{M}_{Edd}$ respectively; and shock parameters: $X_s$ and $R$ 
are varied in between $\sim 25-33$~$r_s$ and $\sim 1.05-1.07$ respectively. Highest value of $\dot{m}_d$ is observed 
on 2017 Sep. 21 (MJD=58017.03), when the highest MAXI/GSC count rate ($5$~Crab) is also recorded. Although depending upon 
the nature of the accretion flow parameters and the QPOs, we have defined outburst period of 2017 Sep. 11 (MJD=58007.27) to 
Oct. 24 (MJD=58050.97) as SIMS, observation of high PL indices ($\Gamma \ge 2.5$) in between 2017 Sep. 21 (MJD=58017.03) to Oct. 11 (MJD=58037.31) 
are quite strange. This high $\Gamma$ period of the outburst could be in SS. As in other objects low frequency QPOs could not be found in SS, 
we defined it as SIMS. Incidentally, Nakahira et al. (2018) and Tao et al. (2018) also reported a similar high $\Gamma$ values. 

\subsection{Mass Estimation}
Mass of the black hole is an important model input parameter while fitting energy spectra with the TCAF model {\it fits} file. 
If mass of the BH is well known, one can freeze mass in that value. However, since mass of MAXI~J1535-571 is not known dynamically, 
to estimate it we keep it as a variable. Each TCAF model fitted spectrum provides us with one best fitted mass, which may 
contain errors due to measurements. As a result we see a fluctuation in {\it derived} mass from one fit to another. 
We find that the model fitted mass during the current outburst of MAXI~J1535-571 varies in a range of $\sim 7.9-9.9$~$M_\odot$ 
(see, Fig. 5a) with an average of mass values of $8.9$~$M_\odot$. So, we may predict probable the mass range of the source from 
our spectral analysis with the TCAF model {\it fits} file as $\sim 7.9-9.9$~$M_\odot$ or $8.9\pm1.0~M_\odot$.

We also estimated mass of the BH using Titarchuk and his collaborators' photon index-QPO frequency 
correlation (scaling) method (Titarchuk \& Fiorito 2004; Shaposhnikov \& Titarchuk 2007, 2009). To do so, we refitted spectral 
data of MAXI J1535-571 using CompTB model (Farinelli et al., 2008) and obtained spectral indices ($\alpha$) for each observations. 
Then in Fig. 6, we plotted the QPO frequency ($\nu$) vs. photon index ($\Gamma=\alpha+1$) for this source (online green circles) 
and 4U 1630-47 during the 1998 outburst (online red squares). Here we use 4U 1630-47 as a reference source since 
they show roughly similar evolution of $\nu$ vs. $\Gamma$ values, which is necessary criteria to use this model. The 4U 1630-47 
data used here are adopted from Seifina, Titarchuk \& Shaposhnikov (2014) paper.
The formula of the scaling method (Shaposhnikov \& Titarchuk 2007) is: 
$$\Gamma(\nu)=A-DB~ln[exp(\frac{\nu_{tr}-\nu}{D})+1], \eqno(1)$$
where A and B are the photon index saturation and slope of the $\Gamma-\nu$ curve. D is a constant which determines how fast 
transition to the saturation of the constant photon indices occurs. $\nu_{tr}$ is the transition frequency above which $\Gamma$ 
saturates. We keep fixed value of $D=1.0$ during our fitting (Shaposhnikov \& Titarchuk 2007). The best-fitted parameters 
are obtained for MAXI J1535-571 (online green curve) as $A_{1535}=2.61\pm0.07$, $B_{1535}=0.24\pm0.03$, $\nu_{tr,1535}=4.9$ 
and for 4U 1630-47 (online red curve) as $A_{1630}=2.66\pm0.02$, $B_{1630}=0.28\pm0.03$, $\nu_{tr,1630}=5$. 
Now, considering 4U 1630-47 as a reference source of mass $9.8\pm0.08~M_\odot$ (Seifina, Titarchuk \& Shaposhnikov 2014), 
we obtain mass of MAXI J1535-571 as, $$M_{1535}=B_{1535}\frac{M_{1630}}{B_{1630}}=8.4\pm1.4~M_\odot,$$
which is well within the range of the TCAF model fitted probable mass values.

\section{Discussion and Concluding Remarks}

The Galactic transient black hole candidate MAXI~J1535-571 was monitored extensively by multi-wavelength band during its 
recent 2017-18 outburst immediately after its discovery on 2017 Sep. 2. We studied evolution of the timing and the spectral 
properties of the source in details during its initial $\sim 50$~days period with the archival or on-demand data of MAXI/GSC, 
Swift/XRT, BAT instruments. Variation of intensities in soft (with MAXI/GSC in $2-10$~keV) and hard (with Swift/BAT 
in $15-50$~keV) X-ray bands and HRs (BAT/GSC count rates) are also studied to find a rough picture about the nature of the outburst. 
Combined XRT ($1-9~keV$), BAT ($15-50$~keV) and/or GSC ($7-20$~keV) or only XRT ($1-9~keV$) spectral data are studied 
during the initial rising phase (from 2017 Sep. 4 to Oct. 24; MJD=58000.71 to 58050.97) of the outburst are studied with 
two types of models: i) phenomenological PL model or DBB plus PL models, ii) Transonic flow solution based physical TCAF model 
{\it fits} file in XSPEC. 

TCAF model fit directly provide us estimation of the physical accretion flow parameters: two types of accretion rates 
(Keplerian disk rate $\dot{m}_d$ and sub-Keplerian halo rate $\dot{m}_h$), and shock parameters (location $X_s$ and compression ratio $R$). 
MAXI~J1535-571 showed rapid evolution of the flow parameters during its initial four observations (MJD=58004-58009). 
As outburst progresses more and more Keplerian matter comes inward and the spectrum becomes softer. 
This phase continues till 2017 Sep. 21 (MJD=58017.03) starting from the 1st day (2017 Sep. 04; MJD=58000.71) of our observation. 
After that the Keplerian disk rate monotonically decreases with time. 
On the first day of our observation, the shock went further ($X_s=269~r_s$) with higher strengths ($R=1.38$), which means that size 
of the Compton cloud or CENBOL became larger. Due to increase in cooling rate, the CENBOL rapidly shrank to a lower radius ($X_s=25~r_s$) 
in between second to fourth observations (2017 Sep. 06-11; MJD=58002-58007). This is caused due to rapid rise in Keplerian disk rate 
and decrease in sub-Keplerian halo rate. After that during the rest of our observation, the shock remained weaker at low radius.
Higher values of ARRs ($> 8.5$) are observed in the first two observations, which is an indication of the object being in the hard spectral state. 
After that in the next two observations, we see a rapid fall in ARR values (from $9.0$ to $0.50$), which is the characteristic signature 
of the hard-intermediate state. After that, during the rest of the observations, ARRs have very low values with slow fluctuations 
in between $0.21-0.63$. This is the direct indication that the source is in the soft-intermediate spectral state.

Low frequency QPOs ($0.44-6.48$~Hz) are observed sporadically on and off during the outburst in $15$~observations out of our 
studied $27$~observations. The detailed nature of the observed QPOs were fitted with Lorentzian profiles and their evolutions 
are given in Table 1 and Fig. 5(c). Based on the nature of the QPOs we see that during the outburst, MAXI~J1535-571 
showed two classical types (A \& C) of QPOs (see, Motta et al. 2011). But the QPO observed on 2017 Sep. 8 (Obs. ID: 00010264003; 
MJD=58004.27) does not fit with any classical type. We marked it as unknown type-X.
According to shock oscillation model (Molteni et al. 1996; Ryu et al. 1997; Chakrabarti et al. 2015), these low frequency QPOs 
are formed due to the resonance oscillation. The QPO frequencies are inversely proportional to the infall time
scales. The shock oscillation occurs when resonance condition satisfies between cooling and infall time scales (Molteni et al. 1996) 
or Rankine-Hugoniot conditions are not satisfy to form a stable shock (Ryu et al. 1997). The sharp type-C QPOs are mainly 
due to resonance oscillation. The lesser sharp, low Q-valued QPOs may occur due to non-satisfaction of Rankine-Hugoniot condition 
or due to weak resonance phenomenon (either a weakly resonating CENBOL for type `B' or a shockless centrifugal barrier for type `A'). 
We should mention that independent of whether the shock actually forms, the flow slows down close to the black hole, but more gradually,
rather than abruptly. So different parts of the barrier will oscillate at a slightly different phase and a slightly different period. 
Combination of so many oscillators will create a bump which looks like a type `A' QPO. This, together with the fact that 
the Rankine-Hugoniot condition is not satisfied (though the flow has three sonic points) makes it possible to have both bump 
and a non-harmonic. The type `A' non-harmonic QPOs observed in Fig. 2a, may be one (low frequency) due to oscillation of the shockless
centrifugal barrier little bit far away from the BH and other one due to non-satisfaction of Rankine-Hugoniot condition close the BH horizon.

Depending upon the nature of the TCAF model fitted/derived flow parameters ($\dot{m}_d$, $\dot{m}_h$, $X_s$, $R$, $ARR$), 
QPOs and PL photon indices, we classified our studied region of the outburst into three spectral states: HS, HIMS and SIMS. 
These are observed in the sequence of,  HS $\rightarrow$ HIMS $\rightarrow$ SIMS. 
Unlike other classical outbursting sources, monotonic evolutions of the QPOs are not observed during the rising HS and HIMS, 
which could be due to non-satisfaction of the resonance condition between cooling and infall time scales (Molteni et al. 1996; 
Chakrabarti et al. 2015). But similar to other BHCs, sporadic signature of QPOs are observed during the rising SIMS. The type-C QPOs 
observed in all three observed spectral states may occur due to resonance oscillation (when cooling time scale roughly matches with 
infall time scales) or unstable shock due to non-satisfaction of Rankine-Hugoniot condition (Molteni et al. 1996; Ryu et al., 1997; 
Chakrabarti et al., 2015). The observed type-A and X QPOs may be due to oscillation of the shockless centrifugal barrier 
or due to non-satisfaction of Rankine-Hugoniot condition. 

In this paper, we only make detailed spectral and temporal study of the source during its very initial phase (only $\sim 50$~days) 
of the outburst starting from 2017 Sep. 4 (two days after its discovery) to Oct. 24. We tried to understand the spectral nature 
of the source during its entire period of the 2017-18 outburst (2017 Sep. 2 to 2018 June 3) from the variation of the HR. 
Here we use the ratio of $15-50$~keV Swift/BAT photon flux with that of the MAXI/GSC in $2-10$~keV band as HR (see, Fig. 1b). 
In BH sources, a fixed energy soft X-ray band photons will not always be emitted from the Keplerian disk and similarly 
fixed energy hard X-ray photons will not always be coming out due to inverse-Comptonization processes. 
Variation of HRs provides us with a guess about the spectral nature of the source during the outburst. 
Depending upon the variation of HRs, we classified the entire outburst into different spectral states. 
Here, similar to other type-I or classical outbursting BHCs (Debnath et al. 2017), we observe four spectral 
states: HS, HIMS, SIMS and SS. These spectral states seem to have followed classical track of 
HS $\rightarrow$ HIMS $\rightarrow$ SIMS $\rightarrow$ SS during the rising phase of the outburst. 

The observation of higher HRs (from $2.83$ to $2.29$) in between 2017 Sep. 5-7 (MJD=58001-58003) support our spectral 
classification of first two observations (on 2017 Sep. 4 \& 6) as HS based on variation of spectral parameters 
and the nature QPOs (if present). Other authors (Nakahira et al. 2018; Tao et al. 2018; Huang et al. 2018; Stiele et al. 2018) 
also reported that during this time the source was in the HS.
Low values of the spectral fitted PL indices ($\Gamma$) ($1.43$ \& $1.45$) also support our conclusion. 
After that, HR is decreased rapidly from $2.29$ to $0.88$ in between 2017 Sep. 7-11 (MJD=58003-58007), which indicates 
that the source is moving towards HIMS. Our result is consistent with Tao et al. (2018) report of HS to HIMS transition 
between 2017 Sep. $8^{th}$ and $11^{th}$. It is also evident from our prediction of 2017 Sep. 8 (MJD=58004.27) observation as HIMS, 
since on this day we observed a sharp fall in ARR ($1.34$ from $9.0$) due to rise in $\dot{m}_d$ and decrease in $\dot{m}_h$. 
After that, HRs are slightly increased or decreased in their lower values till 2017 Nov. 18 (MJD=58075), 
when very low HR (=$0.004$) is observed. We defined this period of the outburst as SIMS, since here sporadic QPOs are also seen. 
Studying the HR diagram Tao et al. (2018) proposed that the source enters in the SIMS
around 2017 Sep. 19 (See also, Huang et al. 2018). Although we see a sudden increase in disk rate ($\dot{m}_d$) and decrease in 
halo rate ($\dot{m}_h$) after 2017 Sep. 18, but from the rapid decrease in shock location ($X_s$), compression ratio ($R$) and
ARR, we suggest the source enters in the SIMS around 2017 Sep. 13.
After that, HRs are remained at its low values till 2018 Apr. 2 (MJD=58210). We guess that during this period of the outburst 
the source was in SS as reported by others (Shidatsu et al., 2017b; Nakahira et al., 2018, Tao et al., 2018). 
On 2018 Apr. 2, high HR=$0.35$ is observed and after this date, we see high fluctuations in HRs from higher to lower values 
(in between $0.01-1.86$). This indicates a evolution of the source towards declining hard or intermediate spectral states. 


Each TCAF model fit provides us one best-fitted mass of the BH when we keep it as 
a free parameter while fitting spectrum. Since the mass of MAXI~J1535-571 is not well known and so far, it has not been measured 
dynamically, we tried to find the probable mass range of the source from our spectral analysis. Model fitted mass values 
are found to vary in the range of $\sim 7.9-9.9~M_\odot$, with an average mass of $8.9$~$M_\odot$ during our analysis period (see, Fig. 5a). 
So, our estimated probable mass range of the source is in between $7.9-9.9~M_\odot$ or $8.9\pm1.0~M_\odot$.
In Fig. 5a, we see a change in the estimated mass of the BH ($M_{BH}$) from observation to observation. This
should not be interpreted as an evolution of mass. In reality, mass  $M_{BH}$ is a derived parameter from our 
spectral fits. So it contains error as in any prediction. Instrumental error and data quality could vary from day to day. 
Also, $M_{BH} \sim T^4$, and thus in TCAF, the spectral fits are highly sensitive to the temperature $T$. So, a small 
error in determination of $T$ of the Keplerian disk gives rise to a significant error in $M_{BH}$. 
We also estimated the mass of the source independently using photon index-QPO frequency correlation (scaling) method 
(Shaposhnikov \& Titarchuk 2007). The estimated mass of MAXI J1535-571 is obtained as $8.4\pm1.4~M_\odot$ when we use 4U 1630-47 
of mass $9.8\pm0.08~M_\odot$ (Seifina, Titarchuk \& Shaposnikov 2014) as a reference source. 
Tao et al. (2018) predicted the mass from diskbb norm as $>14~M_\odot$ for a distance of $10~kpc$ and $>7~M_\odot$ 
for a distance of $5~kpc$ assuming the spin parameter, $a>0.84$ and inclination, $i>57^\circ$.


\section*{Acknowledgment}
This work made use of XRT data supplied by the UK Swift Science Data Centre at the University of Leicester, and MAXI data provided by RIKEN, JAXA and the MAXI team.
J.-R. S., H.-K. C., Y.-X. Y. and J.-L. C. are supported by MOST of Taiwan under grants MOST/106-2923-M-007-002-MY3 and MOST/107-2119-M-007-012.
D.C. and D.D. acknowledge support from DST/SERB sponsored Extra Mural Research project (EMR/2016/003918) fund.
A.J. and D.D. acknowledge support from DST/GITA sponsored India-Taiwan collaborative project (GITA/DST/TWN/P-76/2017) fund.
A.J. also acknowledges CSIR SRF fellowship (09/904(0012)2K18 EMR-1).




\begin{figure}
\begin{center}
\includegraphics[scale=0.6,width=8truecm,angle=0]{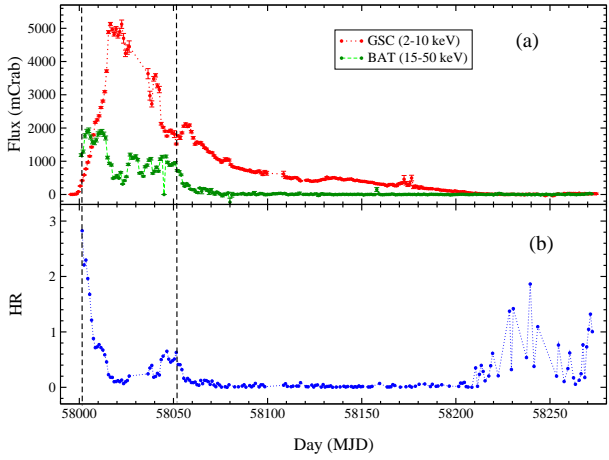}
\caption{(a) Variation of $2-10$~keV MAXI/GSC (dotted-point online red curve) and $15-50$~keV Swift/BAT (dashed-point online green curve) 
one day average fluxes in units of mCrab, 
and (b) variation of the hardness-ratios (HRs=BAT/GSC fluxes) during the 2017-18 outburst of MAXI~J1535-571 are shown.
The period of our spectral analysis presented in the paper are marked by the region in between two vertical lines.
}
\label{fig1}
\end{center}
\end{figure}

\begin{figure}
\begin{center}
\includegraphics[scale=0.6,width=8truecm,angle=0]{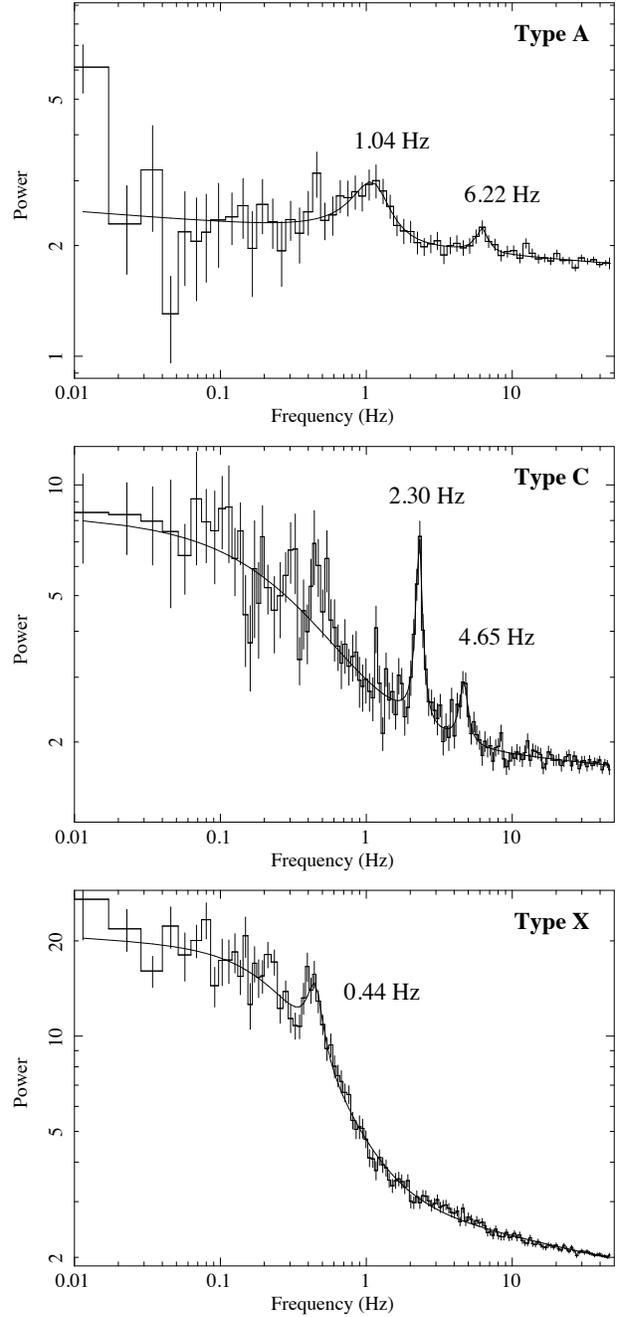}
\caption{(a-c) Lorentzian model fitted Leahy-normalized power density spectrum (PDS) of $0.01$~s time binned XRT light curves in the energy range 
of $1-10$ keV are shown for three observation IDs: 00010264023 (MJD=58029.73), 00010264009 (MJD=58012.11), and 00010264003 (MJD=58004.27). 
QPOs in top two panels belong to the classical type-A or C respectively, while type of the bottom panel QPO is unknown, we marked it as type-X. 
The percentage of rms and Q-values of the primary (high powered) QPOs of the three observations are 
(a) 2.48, 4.70, (b) 4.97, 10.23, and (c) 8.59, 2.34 respectively.}
\label{fig2}
\end{center}
\end{figure}

\begin{figure}
\begin{center}
\includegraphics[scale=0.6,width=8truecm,angle=0]{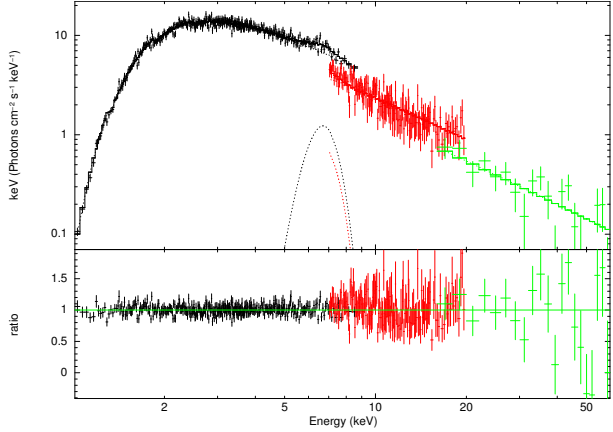}
\caption{TCAF model fitted Swift/XRT (black points), BAT (green points) and MAXI/GSC (red points) combined spectra 
of 2017 Sep. 18 (MJD=58014.17) is shown. The observation ID is 00010264010.}
\label{fig3}
\end{center}
\end{figure}

\begin{figure}
\begin{center}
\includegraphics[scale=0.6,width=8truecm,angle=0]{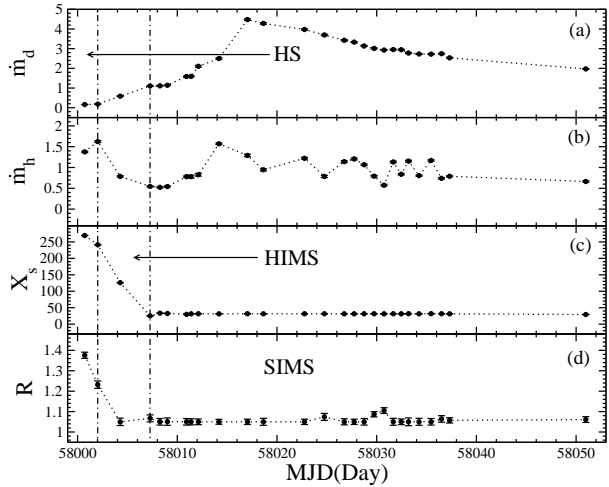}
\caption{(a-b) Variation of the TCAF model fitted flow parameters Keplerian disk rate ($\dot{m}_d$) in $\dot{M}_{Edd}$, 
sub-Keplerian halo rate ($\dot{m}_d$) in $\dot{M}_{Edd}$ are shown in the top two panels. (c-d) Variation of the 
model fitted shock parameters : location $X_s$ in $r_s$ and compression ratio $R$ are shown in the bottom two panels. 
}
\label{fig4}
\end{center}
\end{figure}

\begin{figure}
\begin{center}
\includegraphics[scale=0.6,width=8truecm,angle=0]{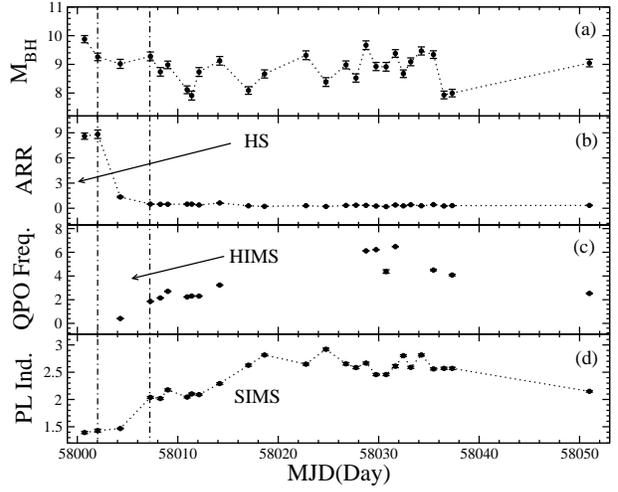}
\caption{Variation of TCAF model (a) fitted mass of the black hole $M_{BH}$ (in $M_\odot$), when it was kept as a free parameter, 
and (b) derived ARR (=$\dot{m}_h$/$\dot{m}_d$) are shown in the top two panels. In the bottom two panels, variation of 
(c) the frequency of the observed QPOs (in Hz), and (d) phenomenological DBB+PL models or with PL model fitted photon indices ($\Gamma$) 
are shown.
}
\label{fig5}
\end{center}
\end{figure}

\begin{figure}
\begin{center}
\includegraphics[scale=0.6,width=8truecm,angle=0]{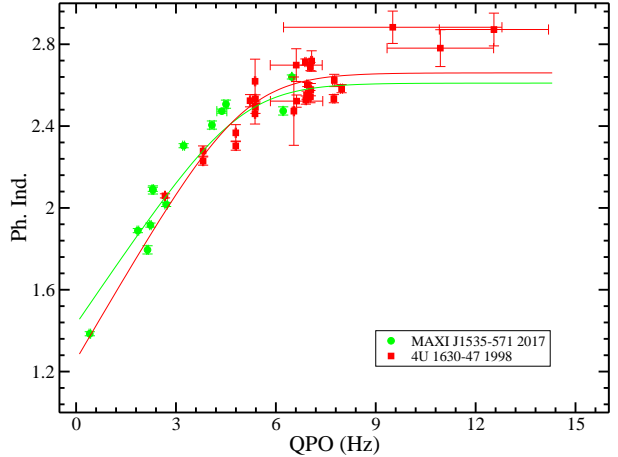}
\caption{Photon index-QPO frequency correlation points during 2017 outburst of MAXI~J1535-571 (online green circle) and 
1998 outburst of 4U~1630-47 (online red square). Fitted curves (online green for MAXI~J1535-571 and red for 4U~1630-47) 
using Eqn. (1) are drawn to find mass of MAXI~J1535-571 with the known (mass) source 4U~1630-47.
}
\label{fig6}
\end{center}
\end{figure}


\clearpage
\begin{table}
\addtolength{\tabcolsep}{-3.0pt}
\centering
\caption{Best-fit Parameters of QPO Fitting Using XRT Data}
\label{tab:table1}
\begin{tabular}{lcccccccc} 
\hline
Obs. ID & Date   &        &  QPO Freq.         & FWHM                 & Q  &  rms  & $\chi^2/dof$ & QPO Type\\
        & (UT) & (MJD)     &  ($\nu$ in Hz)     &  ($\Delta\nu$ in Hz) &    &  (\%) & &\\
 (1)    &  (2) &    (3)    &       (4)          &   (5)                &(6) &  (7)  &(8) &(9)\\
\hline

 00010264003 & 2017-09-08 & 58004.27 & $0.44^{\pm0.01}$ & $0.19^{\pm0.06}$ & $2.3^{\pm0.8} $ & $8.6^{\pm1.9}$ & 146/122 & X\\
 00010264004 & 2017-09-11 & 58007.27 & $1.89^{\pm0.02}$ & $0.25^{\pm0.06}$ & $7.6^{\pm1.9} $ & $5.6^{\pm0.9}$ & 146/116 & C\\
 00010264005 & 2017-09-12 & 58008.26 & $2.15^{\pm0.02}$ & $0.25^{\pm0.07}$ & $8.6^{\pm2.3} $ & $5.8^{\pm1.0}$ & 114/119 & C\\
 00010264007 & 2017-09-13 & 58009.00 & $2.68^{\pm0.03}$ & $0.37^{\pm0.08}$ & $7.2^{\pm1.6} $ & $5.1^{\pm0.8}$ & 138/119 & C\\
 00010264006 & 2017-09-14 & 58010.92 & $2.23^{\pm0.02}$ & $0.29^{\pm0.08}$ & $7.6^{\pm2.2} $ & $5.5^{\pm1.1}$ & 129/119 & C\\
 00010264008 & 2017-09-15 & 58011.38 & $2.31^{\pm0.03}$ & $0.24^{\pm0.06}$ & $9.4^{\pm2.2} $ & $4.8^{\pm0.8}$ & 127/122 & C\\
 00010264009 & 2017-09-16 & 58012.11 & $2.30^{\pm0.02}$ & $0.22^{\pm0.06}$ & $10^{\pm3} $ & $5.0^{\pm0.9}$ & 120/119 & C\\
 00010264010 & 2017-09-18 & 58014.17 & $3.22^{\pm0.03}$ & $0.32^{\pm0.07}$ & $10^{\pm2} $ & $4.2^{\pm0.6}$ & 151/122 & C\\
 00010264022 & 2017-10-02 & 58028.73 & $6.12^{\pm0.19}$ & $0.68^{\pm0.33}$ & $9.0^{\pm4.4} $ & $2.2^{\pm0.7}$ & 166/125 & A\\
 00010264023 & 2017-10-03 & 58029.73 & $6.22^{\pm0.34}$ & $<3.1$ & $<11 $ & $2.5^{\pm1.5}$ & 56/55 & A\\
 00010264024 & 2017-10-04 & 58030.73 & $4.35^{\pm0.04}$ & $0.41^{\pm0.25}$ & $11^{\pm7} $ & $<11.3$ & 71/72 & C\\
 00010264025 & 2017-10-05 & 58031.66 & $6.48^{\pm0.24}$ & $0.82^{\pm0.58}$ & $7.8^{\pm5.5}  $ & $2.5^{\pm1.1}$ & 134/55 & A\\
 00010264029 & 2017-10-09 & 58035.45 & $4.49^{\pm0.05}$ & $0.27^{\pm0.22}$ & $17^{\pm13}$ & $<18.7$ & 168/122 & C\\
 00010264031 & 2017-10-11 & 58037.31 & $4.02^{\pm0.07}$ & $0.58^{\pm0.18}$ & $6.9^{\pm2.1}  $ & $4.5^{\pm0.9}$ & 140/122 & C\\
 00088246001 & 2017-10-24 & 58050.98 & $2.54^{\pm0.04}$ & $0.27^{\pm0.12}$ & $9.4^{\pm4.1}  $ & $4.0^{\pm1.1}$ & 174/119 & C\\

\hline
\end{tabular}
\noindent{
\leftline{Col. 1 represents observation IDs. Col. 2 \& 3 show the dates of observations in UT and MJD, respectively.}
\leftline{Col. 4 \& 5 represent best-fit Lorentzian centroid frequency and full width at half maximum (FWHM) of}
\leftline{observed QPOs in Hz. Col. 6 \& 7 represent Q values (=$\nu/\Delta\nu$) and rms amplitudes of the QPOs.}
\leftline {Note: average values of 90\% confidence $\pm$ error values obtained using `fit err' task, are placed as superscripts of fitted parameters.}
}
\end{table}

\clearpage
\begin{table}
\small
\addtolength{\tabcolsep}{-3.5pt}
\caption{Spectral Fitted Model Parameters}
\label{tab:table2}
\begin{tabular}{lccc|cccccccc}
\hline
Obs. ID$^{[1]}$& \multicolumn{2}{c}{Date} & PL Ind.$^{[2]}$ & $\dot{m}_d$ $^{[3]}$ & $\dot{m}_h$       &        ARR              & $X_s$   & R  &  $M_{BH}$& $N_H$ & $\chi^2/dof$ $^{[4]}$ \\
     & (UT)   & (MJD)    & ($\Gamma$)& ($\dot{M}_{Edd}$) & ($\dot{M}_{Edd}$) & ($\dot{m}_h$/$\dot{m}_d$) & $(r_s)$ &    &($M_\odot$)&  &               \\
 (1) & (2)  & (3) &        (4)       &        (5)        &        (6)              &   (7)   & (8)&   (9)     & (10) &     (11) & (12)       \\
\hline
00770656001*&                  2017-09-04&  58000.71&$1.39^{\pm0.02}$&$0.16^{\pm0.01}$&$1.38^{\pm0.02}$&$  8.6^{\pm0.66}$&$269^{\pm2}$&$ 1.38^{\pm0.02}$&$ 9.9^{\pm0.1}$&$2.35^{\pm0.12}$&335/278 \\
00010264002$^\$$&              2017-09-06&  58002.02&$1.43^{\pm0.03}$&$0.18^{\pm0.01}$&$1.62^{\pm0.02}$&$  9.0^{\pm0.61}$&$241^{\pm3}$&$ 1.23^{\pm0.02}$&$ 9.2^{\pm0.1}$&$3.26^{\pm0.13}$&288/242 \\
00010264003&                   2017-09-08&  58004.27&$1.47^{\pm0.01}$&$0.59^{\pm0.02}$&$0.79^{\pm0.02}$&$ 1.34^{\pm0.08}$&$126^{\pm2}$&$ 1.05^{\pm0.02}$&$ 9.0^{\pm0.2}$&$3.40^{\pm0.12}$&622/606 \\
00010264004*&                  2017-09-11&  58007.27&$2.04^{\pm0.02}$&$1.11^{\pm0.02}$&$0.55^{\pm0.01}$&$ 0.50^{\pm0.02}$&$ 25^{\pm1}$&$ 1.07^{\pm0.02}$&$ 9.3^{\pm0.2}$&$4.42^{\pm0.14}$&621/507 \\
00010264005$^\dagger$&         2017-09-12&  58008.26&$2.02^{\pm0.02}$&$1.11^{\pm0.02}$&$0.52^{\pm0.02}$&$ 0.47^{\pm0.03}$&$ 33^{\pm1}$&$ 1.05^{\pm0.02}$&$ 8.7^{\pm0.2}$&$4.20^{\pm0.13}$&462/437 \\
00010264007*&                  2017-09-13&  58009.00&$2.18^{\pm0.02}$&$1.14^{\pm0.03}$&$0.54^{\pm0.01}$&$ 0.47^{\pm0.02}$&$ 32^{\pm1}$&$ 1.05^{\pm0.02}$&$ 9.0^{\pm0.1}$&$4.01^{\pm0.14}$&406/398 \\
00010264006*&                  2017-09-14&  58010.92&$2.04^{\pm0.02}$&$1.59^{\pm0.02}$&$0.78^{\pm0.02}$&$ 0.49^{\pm0.02}$&$ 30^{\pm1}$&$ 1.05^{\pm0.02}$&$ 8.1^{\pm0.1}$&$4.28^{\pm0.14}$&663/537 \\
00010264008*&                  2017-09-15&  58011.38&$2.10^{\pm0.02}$&$1.59^{\pm0.03}$&$0.78^{\pm0.04}$&$ 0.49^{\pm0.03}$&$ 31^{\pm1}$&$ 1.05^{\pm0.02}$&$ 7.9^{\pm0.2}$&$4.40^{\pm0.14}$&564/517 \\
00010264009*&                  2017-09-16&  58012.11&$2.09^{\pm0.02}$&$2.11^{\pm0.05}$&$0.83^{\pm0.03}$&$ 0.39^{\pm0.02}$&$ 31^{\pm2}$&$ 1.05^{\pm0.01}$&$ 8.7^{\pm0.2}$&$4.61^{\pm0.14}$&586/517 \\
00010264010$^\$$&              2017-09-18&  58014.17&$2.29^{\pm0.02}$&$2.50^{\pm0.04}$&$1.57^{\pm0.02}$&$ 0.63^{\pm0.02}$&$ 31^{\pm1}$&$ 1.05^{\pm0.01}$&$ 9.1^{\pm0.2}$&$4.57^{\pm0.14}$&549/531 \\
00088245003*&                  2017-09-21&  58017.03&$2.63^{\pm0.02}$&$4.48^{\pm0.04}$&$1.29^{\pm0.03}$&$ 0.29^{\pm0.01}$&$ 31^{\pm1}$&$ 1.05^{\pm0.01}$&$ 8.1^{\pm0.1}$&$5.86^{\pm0.14}$&485/442 \\
00010264012*&                  2017-09-22&  58018.63&$2.82^{\pm0.02}$&$4.28^{\pm0.04}$&$0.94^{\pm0.03}$&$ 0.22^{\pm0.01}$&$ 31^{\pm1}$&$ 1.05^{\pm0.02}$&$ 8.7^{\pm0.1}$&$6.11^{\pm0.15}$&673/449 \\
00010264016*&                  2017-09-26&  58022.75&$2.65^{\pm0.02}$&$3.97^{\pm0.04}$&$1.22^{\pm0.02}$&$ 0.31^{\pm0.01}$&$ 31^{\pm1}$&$ 1.05^{\pm0.01}$&$ 9.3^{\pm0.2}$&$5.66^{\pm0.14}$&461/390 \\
00010264018&                   2017-09-28&  58024.75&$2.92^{\pm0.02}$&$3.70^{\pm0.04}$&$0.79^{\pm0.03}$&$ 0.21^{\pm0.01}$&$ 31^{\pm1}$&$ 1.07^{\pm0.02}$&$ 8.4^{\pm0.2}$&$6.29^{\pm0.15}$&451/354 \\
00010264020&                   2017-09-30&  58026.74&$2.65^{\pm0.02}$&$3.43^{\pm0.03}$&$1.14^{\pm0.02}$&$ 0.33^{\pm0.01}$&$ 31^{\pm1}$&$ 1.05^{\pm0.02}$&$ 9.0^{\pm0.1}$&$5.42^{\pm0.14}$&373/343 \\
00010264021&                   2017-10-01&  58027.74&$2.58^{\pm0.02}$&$3.33^{\pm0.03}$&$1.21^{\pm0.02}$&$ 0.36^{\pm0.01}$&$ 31^{\pm2}$&$ 1.05^{\pm0.01}$&$ 8.5^{\pm0.1}$&$5.05^{\pm0.14}$&399/329 \\
00010264022&                   2017-10-02&  58028.73&$2.67^{\pm0.02}$&$3.14^{\pm0.02}$&$1.07^{\pm0.02}$&$ 0.34^{\pm0.01}$&$ 31^{\pm1}$&$ 1.05^{\pm0.02}$&$ 9.7^{\pm0.2}$&$5.02^{\pm0.14}$&461/370 \\
00010264023&                   2017-10-03&  58029.73&$2.46^{\pm0.02}$&$3.01^{\pm0.03}$&$0.79^{\pm0.02}$&$ 0.26^{\pm0.01}$&$ 32^{\pm1}$&$ 1.09^{\pm0.01}$&$ 8.9^{\pm0.1}$&$5.07^{\pm0.14}$&329/352 \\
00010264024&                   2017-10-04&  58030.73&$2.46^{\pm0.02}$&$2.93^{\pm0.02}$&$0.57^{\pm0.02}$&$ 0.19^{\pm0.01}$&$ 31^{\pm1}$&$ 1.10^{\pm0.01}$&$ 8.9^{\pm0.2}$&$5.22^{\pm0.15}$&339/352 \\
00010264025&                   2017-10-05&  58031.66&$2.61^{\pm0.02}$&$2.95^{\pm0.03}$&$1.13^{\pm0.02}$&$ 0.38^{\pm0.01}$&$ 32^{\pm2}$&$ 1.05^{\pm0.02}$&$ 9.4^{\pm0.1}$&$5.40^{\pm0.15}$&233/196 \\
00010264026&                   2017-10-06&  58032.46&$2.80^{\pm0.02}$&$2.95^{\pm0.03}$&$0.84^{\pm0.02}$&$ 0.28^{\pm0.01}$&$ 31^{\pm1}$&$ 1.05^{\pm0.01}$&$ 8.7^{\pm0.1}$&$5.63^{\pm0.15}$&521/387 \\
00010264027&                   2017-10-07&  58033.19&$2.59^{\pm0.02}$&$2.78^{\pm0.02}$&$1.15^{\pm0.02}$&$ 0.41^{\pm0.01}$&$ 32^{\pm1}$&$ 1.05^{\pm0.02}$&$ 9.1^{\pm0.1}$&$5.40^{\pm0.15}$&374/363 \\
00010264028&                   2017-10-08&  58034.24&$2.82^{\pm0.02}$&$2.73^{\pm0.02}$&$0.80^{\pm0.01}$&$ 0.29^{\pm0.01}$&$ 31^{\pm1}$&$ 1.05^{\pm0.02}$&$ 9.5^{\pm0.1}$&$5.26^{\pm0.14}$&577/369 \\
00010264029&                   2017-10-09&  58035.45&$2.56^{\pm0.02}$&$2.72^{\pm0.03}$&$1.17^{\pm0.02}$&$ 0.43^{\pm0.01}$&$ 32^{\pm1}$&$ 1.05^{\pm0.02}$&$ 9.3^{\pm0.1}$&$4.92^{\pm0.14}$&371/364 \\
00010264030&                   2017-10-10&  58036.50&$2.57^{\pm0.02}$&$2.74^{\pm0.02}$&$0.74^{\pm0.02}$&$ 0.27^{\pm0.01}$&$ 32^{\pm2}$&$ 1.06^{\pm0.02}$&$ 7.9^{\pm0.1}$&$5.10^{\pm0.14}$&351/322 \\
00010264031&                   2017-10-11&  58037.31&$2.57^{\pm0.02}$&$2.53^{\pm0.03}$&$0.79^{\pm0.02}$&$ 0.31^{\pm0.01}$&$ 31^{\pm1}$&$ 1.06^{\pm0.01}$&$ 8.0^{\pm0.1}$&$4.99^{\pm0.14}$&351/322 \\
00088246001&                   2017-10-24&  58050.97&$2.15^{\pm0.01}$&$1.97^{\pm0.02}$&$0.66^{\pm0.02}$&$ 0.34^{\pm0.01}$&$ 29^{\pm2}$&$ 1.06^{\pm0.01}$&$ 9.0^{\pm0.1}$&$4.78^{\pm0.14}$&475/440 \\

\hline
\end{tabular}
\leftline{$^{[1]}$ `\$' marks spectra of combined Swift/XRT (1-9 keV), Swift/BAT (15-60 keV) and MAXI/GSC (7-20 keV) data; `*' marks spectra of combined}
\leftline{XRT and BAT data; `$\dagger$' marks spectra of combined XRT and GSC data, and the rests are spectral fitted results only using XRT spectra.}
\leftline{$^{[2]}$ Best-fit photon indices ($\Gamma$) by using the {\it power-law} model are shown in Col. 4.}
\leftline{$^{[3]}$ $\dot{m}_d$, $\dot{m}_h$, $X_s$, $R$, $M_{BH}$ are TCAF parameters. Accretion rates $\dot{m}_d$ and $\dot{m}_h$ are in Eddington rate, and ARR is their ratio ($\dot{m}_h$/$\dot{m}_d$). $X_s$ is the shock}
\leftline{location values in $r_s$ unit. $R$ is the compression ratio and $M_{BH}$ represents the values of mass obtained from the fit in $M_\odot$.}
\leftline{Values of $N_H$ in the unit of $\times10^{22}~atoms~cm^{-2}$ are listed.}
\leftline{$^{[4]}$ Chi-squared values and degrees of freedom of the best fitted TCAF model spectra are presented.}
\leftline {Note: average values of 90\% confidence $\pm$ error values obtained using `err' task in XSPEC, are placed as superscripts of fitted parameters.}
\end{table}

\end{document}